\documentclass{PoS}

\usepackage{graphicx}% Include figure files

\title{Numerical test of the Gribov-Zwanziger scenario in Landau gauge}

\ShortTitle{Test of Gribov-Zwanziger scenario in Landau gauge}

\author{Attilio Cucchieri\\
  Instituto de F\'{\i}sica de S\~ao Carlos, Universidade de S\~ao Paulo, \\ 
  Caixa Postal 369, 13560-970 S\~ao Carlos, SP, Brazil \\
  E-mail: \email{attilio@ifsc.usp.br}}

\author{\speaker{Tereza Mendes}\\%\thanks{A footnote may follow.}\\
  Instituto de F\'{\i}sica de S\~ao Carlos, Universidade de S\~ao Paulo, \\ 
  Caixa Postal 369, 13560-970 S\~ao Carlos, SP, Brazil \\
  E-mail: \email{mendes@ifsc.usp.br}}

\abstract{
 We review the status of lattice simulations of gluon and ghost
 propagators in Landau gauge, testing predictions of the Gribov-Zwanziger
 confinement scenario.
           } 

\FullConference{International Workshop on QCD Green's Functions, 
Confinement, and Phenomenology - QCD-TNT09\\ 
September 07 - 11 2009\\ 
ECT Trento, Italy} 

\def\dblone{\hbox{$1\hskip -1.2pt\vrule depth 0pt height 1.6ex width 0.7pt
\vrule depth 0pt height 0.3pt width 0.12em$}}
 
\begin{document} 

\section{Introduction} 

The investigation of infrared properties of Landau-gauge gluon and ghost 
propagators --- in order to test predictions of the Gribov-Zwanziger
confinement scenario, or of the Kugo-Ojima scenario --- has generated a
flurry of papers in the last couple of years.
While the formulation of these two scenarios, their equivalence (or
lack thereof) and the chosen procedure for solving these propagators' 
Dyson-Schwinger equations have become matters of heated debate, 
a consistent picture emerges from lattice studies. Based on this picture, 
which is in partial disagreement with the predictions 
of both scenarios, present activity focuses on critical revisions of 
the original scenarios, on discussion of their main assumptions and/or 
implications, and on whether or not there are physical criteria to
prefer a solution of the Dyson-Schwinger equations (or other functional 
methods) of the ``scaling'' or of the ``massive'' type. 
In this way, one hopes to have gained a 
deeper insight into confinement in Landau gauge. We will not review
these various analytic studies here, but rather refer to recent 
status reports and overviews contained in these proceedings, such
as Ref.\ \cite{Dudal:2009bf}, which summarizes the so-called refined
Gribov-Zwanziger framework, Refs.\ \cite{Binosi:2009er} and 
\cite{Kondo:2009qp}, which address the problem of the characterization
of the Kugo-Ojima scenario and its relation to the Gribov-Zwanziger
one, or Ref.\ \cite{Natale:2009uz}, which reviews the solution of
Dyson-Schwinger equations in Landau gauge and discusses phenomenological
applications. 
We do, however, attempt to review thoroughly the recent literature 
on lattice studies of the topic.

In what follows we consider tests of the original Gribov-Zwanziger 
confinement scenario, i.e.\ a vanishing gluon propagator and an 
enhanced ghost propagator in the infrared limit. On the contrary,
in the refined Gribov-Zwanziger framework mentioned above, one expects 
a finite (nonzero) gluon propagator and a free ghost propagator in the 
same limit. These two cases are consistent respectively with the so-called 
scaling and massive (or decoupling) solutions of the Dyson-Schwinger 
equations.
The latter behavior has been strongly favored in all recent 
extensive lattice studies, performed on very large lattices
for pure SU(2) and SU(3) gauge theory. 
As argued in \cite{Cucchieri:2008yp}, the essential features of the 
original Gribov-Zwanziger confinement scenario are not incompatible
with these findings, since violation of reflection positivity is
clearly observed for the gluon propagator and enhancement with respect
to the free propagator is seen for the ghost at intermediate momenta.
Also, a logarithmic enhancement of the ghost propagator might be possible 
in the continuum limit.
A scaling solution (with nontrivial infrared exponents), however, 
is ruled out.

From the point of view of lattice simulations, we must strive to keep
under control the various sources of systematic errors that might obscure
the true infrared behavior of the propagators, in order to conclude that
the behavior described above is firmly established.
Of course, once the massive behavior is confirmed from the simulations, 
one must understand why this behavior arises in the infrared limit of the
theory, and how we might reconcile it with a confinement mechanism. Again, 
insight into the problem can hopefully come from the numerical simulations
themselves. Also, it should be noted that the comparisons mentioned above
assume that the gauge definition on the lattice is physically equivalent 
to the continuum one, an issue that should be carefully investigated. 
We will refer here to the so-called minimal Landau gauge condition
\cite{Zwanziger:1993dh}.

In Section \ref{syst} we review some aspects of the numerical simulations,
with special attention to the main possible sources of systematic errors. 
We also attempt an overview of references and chronology of the main recent 
lattice results on the topic. 
In Section \ref{bounds} we summarize interesting constraints
on the infrared behavior of the propagators, written in the form
of upper and lower bounds at fixed lattice volume. Although these
bounds were introduced as a guide to the infinite-volume extrapolation,
we believe they will be useful tools to investigate why the
propagators have the observed behavior, since they naturally relate to
a statistical interpretation of the gluon propagator and to a
clearer view of the ghost propagator in terms of the spectrum of
the Faddeev-Popov operator. Section \ref{huge} is dedicated to the
results from very large lattices mentioned above. We summarize the
analysis of data from our simulations of the pure-SU(2) case, which 
are essentially equivalent to the corresponding results by other
lattice groups.
In Section \ref{beta0} we address the $\beta=0$ case, where various 
sources of systematic errors may be investigated more easily. 
The observed behavior is compared with the one at finite $\beta$. 
Finally, we present our conclusions in Section \ref{conclusion}.

\section{The simulations}
\label{syst}

On the lattice, the gauge action is written in terms of oriented 
plaquettes, formed by the link variables $U_{\mu}(x)$, which are
elements of the gauge group $SU(N_c)$ and change under gauge 
transformations as 
\begin{equation}
U_{\mu}(x)\;\to\; U_{\mu}^g(x) \;\equiv \;
g(x)\,U_{\mu}(x)\,g(x+{\hat \mu})^{\dagger}\,,
\label{eq:Ug}
\end{equation} 
where $g\in SU(N_c)$, lattice sites are labeled by $x$ and $\mu$
refers to the directions along the lattice.
Consequently, all closed loops are gauge-invariant 
quantities, including the plaquettes in the pure-gauge-theory action
\begin{equation}
S(\{U\}) \;=\; \frac{\beta}{N_c} \,\sum_{\mu<\nu}\sum_x
\Re\,{\rm Tr} \left[ \dblone \,-\, 
U_{\mu}(x) U_{\nu}(x+{\hat \mu}) 
U_{\mu}^{\dagger}(x+{\hat \nu}) U_{\nu}^{\dagger}(x)\right]\,.
\end{equation}
Here $\beta$ is the lattice parameter, related to the bare 
coupling constant $g_0^2\,$ as $\,\beta = 2 N_c/g_0^2$.
Except where otherwise indicated, we consider (symmetric) 
hypercubic lattices of four space-time dimensions.
The lattice size in physical units is given by $L= N a$, where $N$
is the number of points per lattice direction and the lattice
spacing $a$ is expressed in physical units. The physical volume is
thus $V = L^4$.
(Note that one often refers to the ``lattice volume'' $N^4 = V/a^4$.)
We assume periodic boundary conditions.

Let us remark that the resulting path integral has a finite (group) 
integration volume and there is in principle no need for gauge fixing 
on the lattice.
A procedure for fixing the gauge numerically may nevertheless be 
implemented in a straightforward way in the simulation, without the need 
to consider the Faddeev-Popov matrix ${\cal M}$. For minimal Landau gauge, 
one minimizes the functional
\begin{equation}
{\cal E}(\{g\})\;=\;
1\,-\, \frac{a^4}{4\,N_c\,V}\sum_{\mu}\sum_x \Re\,{\rm Tr}\,U^g_{\mu}(x)
\end{equation}
with respect to the gauge transformations $\{g\}$.
Indeed, the first derivative of this functional yields the
familiar Landau gauge condition of null (lattice) divergence of
the gauge field, which is defined in terms of the link variables as
\begin{equation}
A_{\mu}(x) \;=\; \frac{1}{2 i\, a g_0} 
\left[U_{\mu}(x)-U_{\mu}^{\dagger}(x) \right]_{\rm traceless} \,.
\end{equation}
The prescription for fixing the gauge in a simulation is thus to:
1) produce a gauge-link configuration $\{U\}$ as usual,
2) to find $\{g\}$ that is a (local) minimum of the functional 
${\cal E}$ (holding $\{U\}$ fixed) and 3) to transform $\{U\}$ following 
Eq.\ (\ref{eq:Ug}) for the selected $\{g\}$. The resulting configuration 
is a gauge-fixed realization of the link variables, which will be used
to compute observables of interest such as the gluon and ghost propagators.

The gluon propagator is given in Landau gauge simply by
\begin{equation}
D^{ab}_{\mu \nu}(p) \;=\; \sum_x  e^{-2 i \pi p \cdot x}
\langle A^a_{\mu}(x)\,A^b_{\nu}(0) \rangle \;=\;
\,\delta^{ab}\,\left(g_{\mu \nu}\,-\,\frac{p_{\mu}\,p_{\nu}}{p^2}\right)
D(p^2)\,,
\end{equation}
where $p$ is the momentum and $a$, $b$ are color indices.
It is therefore determined solely by the scalar function $D(q^2)$
associated to its transverse component.
In the original Gribov-Zwanziger scenario, gluon confinement is
associated with violation of reflection positivity for the gluon
propagator (in real space-time).

As pointed out above, lattice gauge fixing is accomplished without
the need to compute the Faddeev-Popov matrix ${\cal M}$. Nevertheless,
the matrix can be obtained directly from the second variation of the 
gauge-fixing functional ${\cal E}$, which corresponds to the
Jacobian of the gauge-fixing condition. It is interesting to note that
in this way there is also no need to consider the ghost fields
explicitly \cite{Mandula:1999nj}. The ghost propagator $G(p^2)$ is 
given by the inverse of ${\cal M}$ as
\begin{equation}
G(p^2)\; = \;\frac{1}{N_c^2 - 1} \, \sum_{x\mbox{,}\, y\mbox{,}\, a}
\frac{e^{- 2 \pi i \, p \cdot (x - y)}}{V}\,
\langle\, {\cal M}^{- 1}(a,x;a,y) \,\rangle\,.
\end{equation}
An infrared enhancement of $G(p^2)$ with respect to the tree-level ghost 
propagator $G(p^2)\sim p^{-2}$ is expected in the original Gribov-Zwanziger
scenario (and in the Kugo-Ojima one) as a sign of confinement.

Let us note that the known problem of Gribov copies is present on the 
lattice as well, since each local minimum of the functional ${\cal E}$
corresponds to an equivalent (lattice) gauge copy. The algorithm 
for fixing $\{g\}$ has in principle no control over which copy gets 
selected. Because of the minimization, we know that ${\cal M}$ is
positive semi-definite and, as a result, the sampled copies are inside 
the first Gribov horizon $\Omega$, which is delimited by the vanishing of 
$\lambda_{min}$, the smallest nontrivial eigenvalue of ${\cal M}$. 
It is usually argued that a unique copy might be obtained, corresponding 
to the fundamental modular region $\Lambda$, if one were able to determine 
the global minimum of ${\cal E}$.
This region has been studied on the lattice in \cite{Cucchieri:1997ns}.
In any case, $\Omega$ and $\Lambda$ are shown to be convex regions of 
very high dimensionality, which likely constrains the statistical 
weight of gauge configurations to lie near their boundary. In particular, 
one should check if the sampled configurations have vanishing
$\lambda_{min}$ as the lattice volume goes to infinity.

Lattice simulations have been carried out since the mid 1980s for 
the gluon propagator \cite{Mandula:1987rh} and since the mid 1990s 
for the ghost \cite{Suman:1995zg} (see also \cite{Cucchieri:1997dx}).
Early studies have established that the gluon propagator is not
enhanced at small momenta \cite{Leinweber:1998uu}, but did not allow 
further conclusions about its infrared behavior. A turnover point in 
momentum (suggesting a null propagator) could only be seen at strong
coupling \cite{Cucchieri:1997fy}
or in three space-time dimensions \cite{Cucchieri:1999sz}. 
However, even in this simplified case, a later study on a very large 
lattice (of volume $140^3$) was still not conclusive 
\cite{Cucchieri:2003di}, although it was possible to fit $D(0)$ to 
zero in an infinite-volume extrapolation, and violation of
reflection positivity was clearly seen \cite{Cucchieri:2004mf}.
For the ghost propagator, infrared enhancement was observed,
but the corresponding infrared exponent seemed to become smaller 
as lower momenta became available \cite{Suman:1995zg,Cucchieri:1997dx,
ghost,%Furui:2003jr,Gattnar:2004bf,Sternbeck:2005tk,
Cucchieri:2006tf}.
It was also clearly shown that $\lambda_{min}$ goes to zero with 
increasing lattice volume \cite{Sternbeck:2005vs,Cucchieri:2006tf}.
These studies were complemented by investigations of the strong
coupling constant (see e.g.\ \cite{Boucaud:1998bq,Bloch:2003sk})
and several three-point vertices (see e.g.\ 
\cite{Parrinello:1994wd,Skullerud:2002ge,Cucchieri:2004sq,Cucchieri:2006tf}).
This was the status until 2006.
As became clear later, systematic effects had not yet been properly taken
into account, which limited the conclusions (or the lack thereof) of 
these studies. We now pause for a moment and list the main such 
possible effects (and related references) below.
\begin{itemize}
\item{\bf Gribov-copy effects}: this is a very important issue.
As commented above, usual simulations do not take fluctuations
in the values of the propagators due to Gribov-copy effects into 
account. 
A few studies have considered the determination of the absolute 
minimum of the gauge-fixing functional 
(see e.g.\ \cite{Cucchieri:1997dx,Gribovcopies}) %Silva:2004bv,Maas:2008ri})
or other criteria to fix the gauge \cite{absmin}.
%Bogolubsky:2005wf,Maas:2009se}.
It was generally found that the effect of Gribov copies decreases 
as the lattice volume increases.
This statement must be taken with a grain of salt, since the number 
of copies surveyed is limited and one does not know for sure if this 
number is large enough at a given volume to allow the determination
of the global minimum. Studies of the exact structure 
of Gribov copies are now being carried out (on small lattices)
\cite{Mehta:2009zv}. 
Let us mention that it was argued by Zwanziger \cite{Zwanziger:2003cf} 
that averages taken in the fundamental modular region should coincide 
with averages in $\Omega$ in the infinite-volume limit. 
We thus conclude that a sign of significant Gribov-copy effects has not 
yet been seen and the effects observed so far are probably connected to 
the next item below. 
(We do note, however, that a very recent study has reported on sizeable
effects at large lattice volumes \cite{Bogolubsky:2009qb}.)

\item{\bf Finite-volume effects}: perhaps surprisingly, these are the
most serious systematic effects we have to deal with. To be sure, 
lattice simulations must be carried out at finite lattice volumes, 
since computers have finite memory. As mentioned above, the physical extent
of the lattice $L$ is given by the number of lattice points along each 
direction multiplied by the lattice spacing $a$ in physical units, which 
is directly related to the lattice parameter $\beta$. 
To simulate closer to the continuum limit 
one must go to smaller $a$, or equivalently to larger $\beta$, 
while keeping the simulated lattice large enough to represent the
relevant energy scales of the problem. Strictly speaking, 
one would need an extrapolation to infinite lattice volume at 
each fixed value of $\beta$. (A continuum 
extrapolation would additionally require running at increasingly smaller 
values of $a$.) In usual lattice applications, though, taking the 
infinite-volume limit is not among the most serious issues, since one 
typically just needs to have a sufficient number of points to ensure
a physical lattice size of the order of the relevant hadronic scale, 
i.e.\ around 1 fm. 
The main effort is then to go to very small $a$, in order to avoid 
discretization errors (addressed below). In studies of the infrared limit, 
however, the situation is different, and finite-size effects play an 
important role. This happens because the infrared limit lies at small 
$p$, corresponding to large $L$. (Note that the smallest nonzero momentum 
that can be represented on a lattice of side $L$ is $\sim 2\,\pi /L$.)

\item{\bf Discretization effects}: as indicated above, the effects due
to simulating at nonzero lattice spacing $a$ are not expected to be so
serious in the infrared limit, because the energy scale associated with
the cutoff $a$ (which is $\sim 1/a$), is sufficiently high compared to
the typical momenta of interest. In other words, the long wavelengths 
we are interested in do not resolve the lattice spacing and are not 
much affected by it. Nevertheless, discretization errors may be 
important for the breaking of rotational symmetry as well as for 
the possible different discretizations of the gluon field and of the
gauge-fixing condition. There are ways to reduce effects due to the 
breaking of rotational symmetry, such as cutting out the momenta
characterized by large effects \cite{Leinweber:1998uu} (the so-called 
cylindrical cut), improving the lattice definition of the momenta 
\cite{Ma:1999kn} and including (hypercubic) corrections into the 
momentum-dependence of the Green's functions 
\cite{deSoto:2007ht}. As for the discretization of the gluon field and 
of the lattice Landau gauge condition, several different definitions
may be considered (see e.g.\ \cite{discretization}).
%\cite{Giusti:1998ur,Bonnet:1999mj,Bloch:2003sk,vonSmekal:2007ns}).
These studies have 
usually found that different discretization procedures lead to gluon 
propagators that differ only by a multiplicative constant, which can
be reabsorbed in the (multiplicative) renormalization of the propagator.

\item{\bf Unquenching}: in the Gribov-Zwanziger and related scenarios, 
one hopes to get an understanding of confinement in the static-quark
limit, where there is no string breaking and the confinement problem
may be phrased as a search for explaining why an area law develops
\cite{Greensite:2003bk}. Thus, it should be sufficient to consider the 
pure-gauge theory treated here, also known as the quenched approximation. 
Nevertheless, an important question is how the picture gets affected 
once dynamical quarks are introduced in the simulations.
Studies done so far (on relatively small lattices) show qualitatively
the same behavior as in the pure-gauge case \cite{unquenched}.
%\cite{Boucaud:2001un,Furui:2005bu,Ilgenfritz:2006he,Bowman:2007du}.

\end{itemize}

In 2007, studies of Dyson-Schwinger equations on the torus 
\cite{Fischer:2007pf} hinted that finite-size effects might indeed 
be plaguing results from lattice simulations and predicted that 
physical lattice sides of the order of 15 fm might be needed
to begin to see the expected (conformal scaling) infrared behavior 
of the propagators. 
At about the same time, two other predictions of similar studies
were verified in simulations: the (quantitative)
equivalence of infrared propagators in the SU(2) and SU(3) 
cases\footnote{
Although one cannot expect this to hold at high values of $p$, 
it is conjectured that the cases $N_c=2$ and $N_c=3$ have the 
same infrared behavior. A recent comparison along a wider 
range of momenta presented in \cite{Oliveira:2009nn}
shows some discrepancies between the two cases.}
\cite{Cucchieri:2007zm,Sternbeck:2007ug}
and the verification of conformal scaling behavior in two space-time
dimensions \cite{Maas:2007uv}.
That same year, three groups came out with studies on very large
lattices, which were all presented at the {\em Lattice 2007}
conference. The Berlin-Dubna group considered $80^4$ lattices, corresponding
to a lattice extent of 13 fm, in the SU(3) case \cite{Bogolubsky:2007ud}.
(Their study of the gluon propagator was later extended to $96^4$ lattices, 
corresponding to a lattice extent of 16 fm \cite{Bogolubsky:2009dc}.)
The Adelaide group considered $112^4$ lattices, corresponding
to 19 fm, in the SU(2) case \cite{Sternbeck:2007ug}.
Similarly, we considered $128^4$ lattices (corresponding to 27 fm)
in the SU(2) case, plus three-dimensional lattices of size
$320^3$, corresponding to 85 fm \cite{Cucchieri:2007md}.
What these studies showed was puzzling. On the one hand, the large
volumes clearly allowed a better view of the infrared picture; on
the other, this view was nothing like what the authors had imagined it 
would be!
In fact, going to large volumes in the hopes of seeing a null infrared 
gluon propagator not only established that $D(0)$ was {\em not} null 
(even in three dimensions), but also exposed
the fact that the previously seen enhancement of the ghost propagator 
goes away at very small momenta, and the data are consistent with a 
flat ghost dressing function in the infrared limit.

This behavior is analyzed in Section \ref{huge}. In the next section,
on the other hand, we comment on rigorous bounds for the propagators,
introduced as a guide to the infinite-volume extrapolation.

\section{Bounds on propagators and statistical interpretation}
\label{bounds}

   As discussed in the previous section, one of the main difficulties in
the lattice simulations is the extrapolation of gluon- and ghost-propagator 
data to infinite lattice volume. In fact, the correct volume dependence 
of the data may not be easily inferred from the behavior on medium-size 
(or even very large) lattices, especially since some quantities, such as 
the zero-momentum gluon propagator, are quite noisy. It is then very
helpful to obtain constraints on the infrared behavior of the propagators, 
as the upper and lower bounds discussed in this section. We remark that 
these bounds are valid at each lattice volume and must be extrapolated 
to infinite volume, just as for the propagators. The advantage is that
the bounds are written in terms of quantities that are more intuitive 
than the propagators themselves, making it easier to guess the
expected volume dependence of the propagators and possibly allowing an
explanation of the infrared behavior observed in the data.

In the case of the gluon propagator, we obtain \cite{Cucchieri:2007rg}
the bounds
\begin{equation}
V \, {\langle {M}(0) \rangle}^2 \, \leq \; D(0)\;
    \leq \; V d (N_c^2 - 1) \, \langle {{M}(0)}^2 \rangle \; ,
\label{eq:Dbounds}
\end{equation}
where $d$ is the dimension, $V$ is the volume,
\begin{equation}
D(0) \; = \; \frac{V}{d (N_c^2 - 1)} \sum_{\mu, b}
\langle | {\widetilde A}^b_{\mu}(0) |^2 \rangle
\end{equation}
is the zero-momentum propagator and $M(0)$ is defined as
\begin{equation}
{M}(0) \, = \, \frac{1}{d (N_c^2 - 1)}
   \sum_{b,\mu} | {\widetilde A}^b_{\mu}(0) | \; .
\label{eq:mag}
\end{equation}
Let us also define the ``magnetization''
\begin{equation}
{M}'(0) \, = \, \frac{1}{d (N_c^2 - 1)}
   \sum_{b,\mu} {\widetilde A}^b_{\mu}(0) \; .
\end{equation}
From the above definitions we can get a {\em statistical interpretation}
for the quantity on the right-hand side of (\ref{eq:Dbounds}): it is
essentially the susceptibility associated with the magnetization ${M}'(0)$
(since the average of this magnetization vanishes, due to the residual
global gauge symmetry). By analogy with a $d$-dimensional spin system 
one would thus expect to see $ \,V \langle {{M}(0)}^2  \rangle \sim const$, 
i.e.\ the statistical variance of the magnetization is proportional to the 
inverse of the volume, a behavior known as {\em self-averaging}.
At the same time, considering the statistical fluctuations in the Monte Carlo
sampling of ${M}(0)$, we would expect ${\langle {M}(0) \rangle}^2 $ to have
the same volume dependence as $\langle {{M}(0)}^2 \rangle$ 
\cite{Cucchieri:2007rg}.
The simple statistical argument presented above suggests that both
${\langle {M}(0) \rangle}^2$ and $\langle {{M}(0)}^2 \rangle$ should show a
volume dependence as $1/V$. On the other hand, this suppression with $1/V$ is
compensated by the volume factor for both bounds in Eq.\ (\ref{eq:Dbounds}).
Consequently, if this suggested behavior for the susceptibilities is verified,
$D(0)$ converges to a nonzero constant in the infinite-volume limit.
As explained in the next section, this is what one observes in the
simulations.
Note that the bounds in Eq.\ (\ref{eq:Dbounds}) apply to any gauge and
that they can be immediately extended to the case $D(p^2)$ with $p \neq 0$.
We also note that a very interesting stochastic interpretation for the
gluon propagator has been investigated in \cite{Suganuma:2009zs}.

Also in the case of the ghost propagator, a more intuitive picture 
comes from noticing that in Landau gauge, for any nonzero momentum $p$, 
one finds \cite{Cucchieri:2008fc}
\begin{equation}
\frac{1}{N_c^2 - 1} \,  \frac{1}{\lambda_{min}} \, \sum_a \,
  | {\widetilde \psi_{min}(a,p)} |^2 \,
              \leq \, G(p^2) \, \leq \, \frac{1}{\lambda_{min}} \; ,
\label{eq:Gineq}
\end{equation}
where $\lambda_{min}$ is the smallest nonzero eigenvalue of the Faddeev-Popov
operator ${\cal M}$ and ${\widetilde \psi_{min}(a,p)}$ is the corresponding
eigenvector. Note that the upper bound is independent of the momentum $p$.
If we now assume $\lambda_{min}\sim L^{-\nu}$ and
$\,G(p^2) \sim p^{-2-2\kappa}$ at small $p$, we have that
$2+2\kappa \leq \nu$, i.e.\ $\nu > 2$, is a necessary condition for
the infrared enhancement of $G(p^2)$. A similar analysis can be carried
out for a generic gauge condition.
Consider the Gribov region $\Omega$, where all eigenvalues
of ${\cal M}$ are positive. In the infinite-volume limit,
as mentioned above, entropy favors configurations near the Gribov 
horizon $\partial \Omega$, where $\lambda_{min}$ goes to zero. 
Thus, inequalities such as (\ref{eq:Gineq}) can tell us if one should 
expect an enhanced $G(p^2)$ when the Boltzmann weight gets concentrated 
on $\partial \Omega \,$.
(This answers the question posed in \cite{Cucchieri:2006hi}.)
In other words, it is clearly necessary --- but not sufficient --- to 
have a vanishing $\lambda_{min}$ as the volume tends to infinity in order
to observe enhancement of $G(p^2)$. 
The upper bound in Eq.\ (\ref{eq:Gineq}) was tested for our data in 
\cite{Cucchieri:2008fc}, albeit with limited statistics for 
$\lambda_{min}$. 
We find $\nu\approx 2$, consistent with finding $\kappa\to 0$ from
fits of $G(p^2)$, as shown in the next section.

Let us also mention that a possible connection between the infrared behavior 
of the gluon propagator and the appearance of nontrivial zero modes of 
${\cal M}$ for configurations near $\partial \Omega$ has been recently
presented in \cite{Greensite:2010hn}.
The two types of scenarios obtained there can probably be related to
the massive and the scaling solution for the propagators.

\section{Huge lattices}
\label{huge}
\begin{figure}[t]
\vspace{-22mm}
\hspace{-4mm}
\includegraphics[scale=0.44]{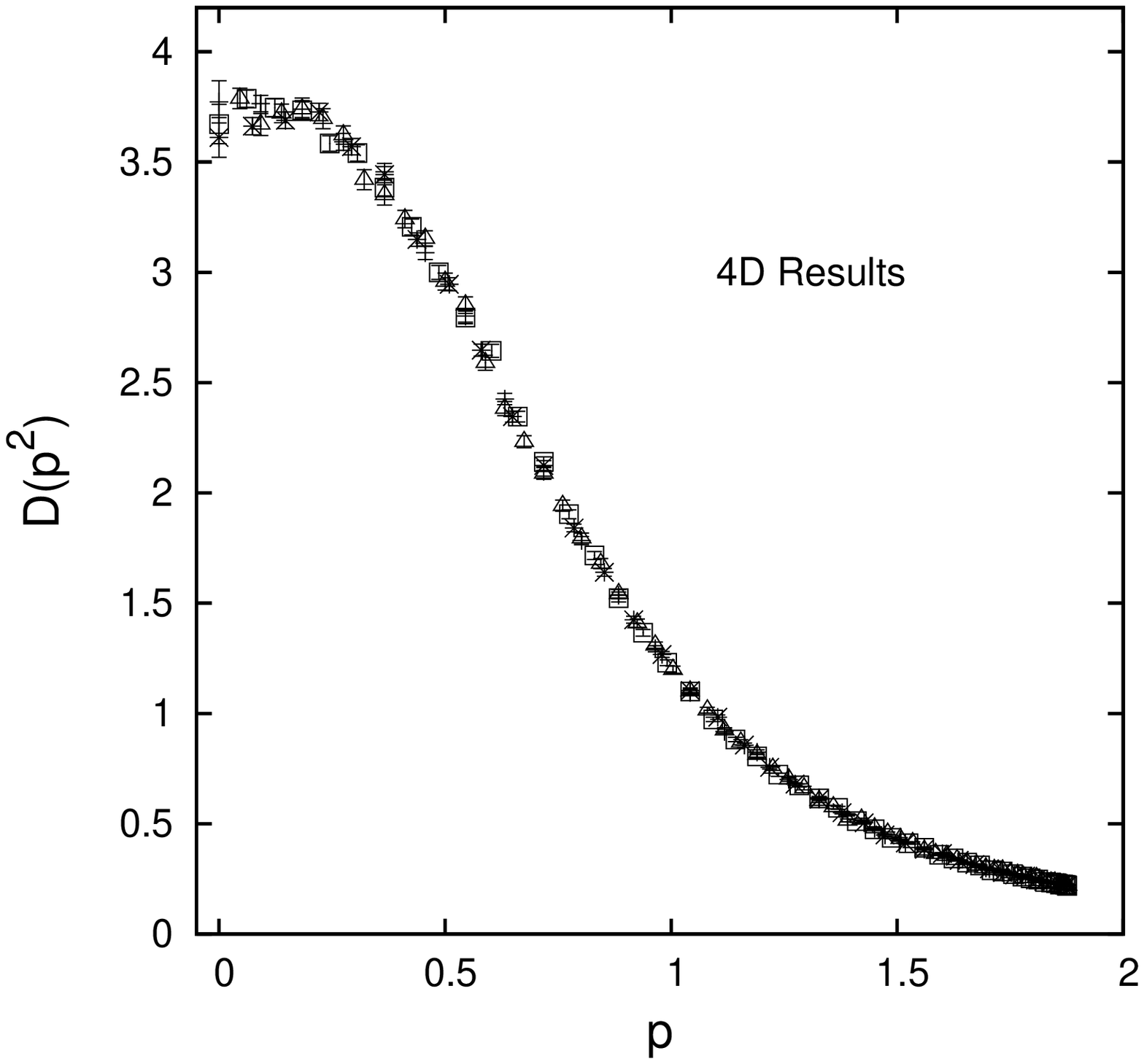}
\hspace{-4mm}
\includegraphics[scale=0.44]{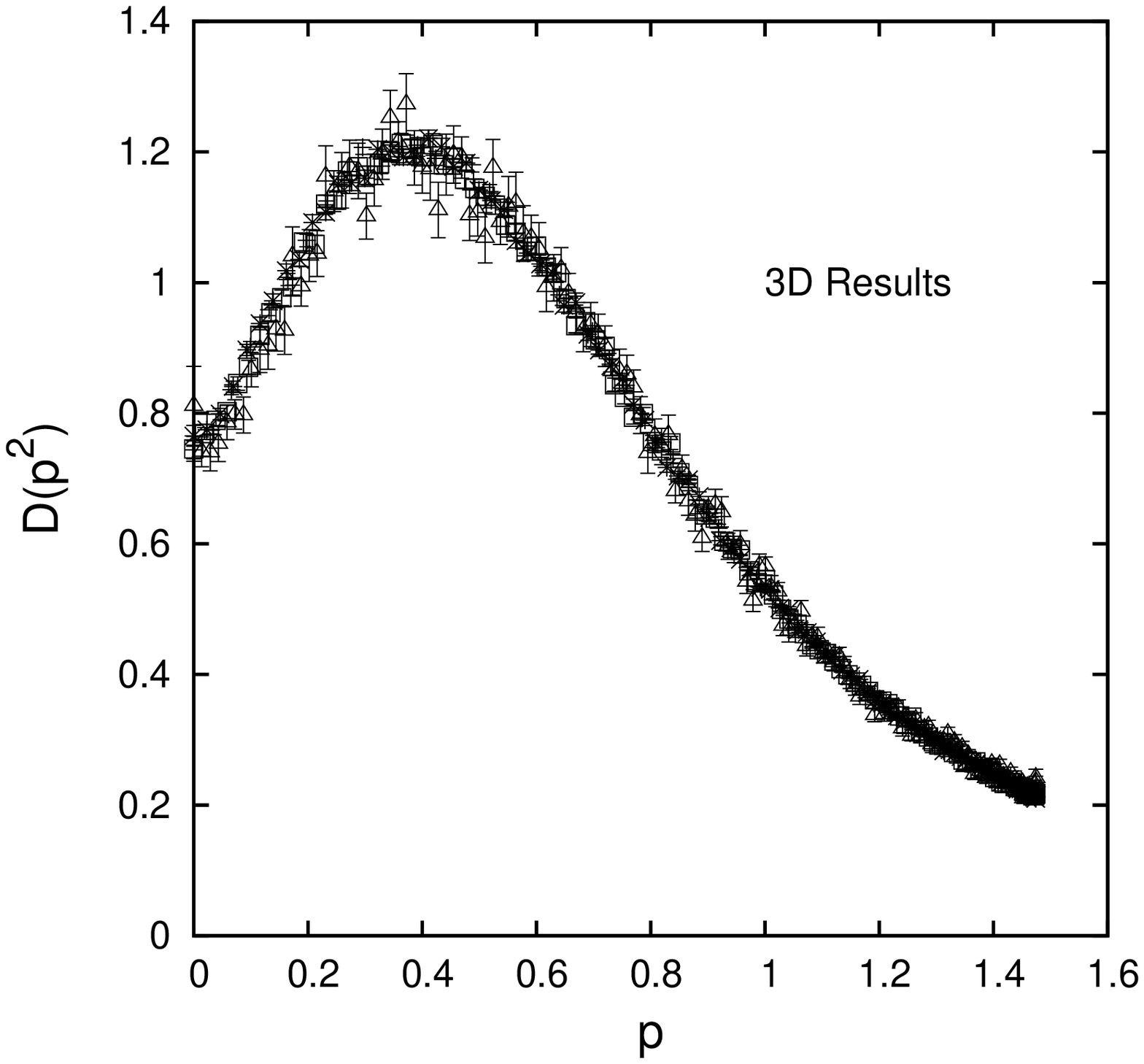}
\vspace{-26mm}
\caption{
  The gluon propagator $D(p^2)$ as a function of the
  lattice momenta $p$ (both in physical units) for the pure-$SU(2)$ 
  case in $d=4$ (left), considering volumes of up to $128^4$ 
  (lattice extent $\sim 27$ fm), and $d=3$ (right),
  considering volumes of up to $320^3$ 
  (lattice extent $\sim 85$ fm).
  }
\label{gluon}
\end{figure}
Our study in the $SU(2)$ Landau case \cite{Cucchieri:2008fc}, 
using the very large lattices mentioned above, is summarized here.
In Fig.\ \ref{gluon} we show data for the gluon propagator in $d=4$
and $d=3$ for a wide range of (large) lattice volumes, indicating
that $D(0)$ remains nonzero in the infinite-volume limit. 
Similar results are obtained in 
\cite{Sternbeck:2007ug,Bogolubsky:2007ud,Bogolubsky:2009dc}, as
mentioned before, but also in \cite{Bornyakov:2008yx},
which takes Gribov-copy effects into account and in 
\cite{Gong:2008td}, which uses improved actions and anisotropic lattices.
We have investigated the volume dependence of the bounds in 
Eq.\ (\ref{eq:Dbounds})
and found remarkably good agreement with the predicted $1/V$ behavior 
for ${\langle {M}(0) \rangle}^2$ and $\langle {{M}(0)}^2 \rangle$,
thus implying a finite nonzero value for $D(0)$ in the infinite-volume limit.
More precisely, by fitting the two quantities to $1/V^{\alpha}$ we get 
the exponents $\alpha$ respectively 0.995(10) and 0.998(10). A similar
analysis for the SU(3) case (considering somewhat smaller volumes) yields 
the exponents 1.058(6) and 1.056(6) \cite{Oliveira:2008uf}.
Violation of reflection positivity for $D(x)$ is seen in all cases.

\begin{figure}[t]
\vspace{-17mm}
\hspace{-3mm}
\includegraphics[scale=0.43]{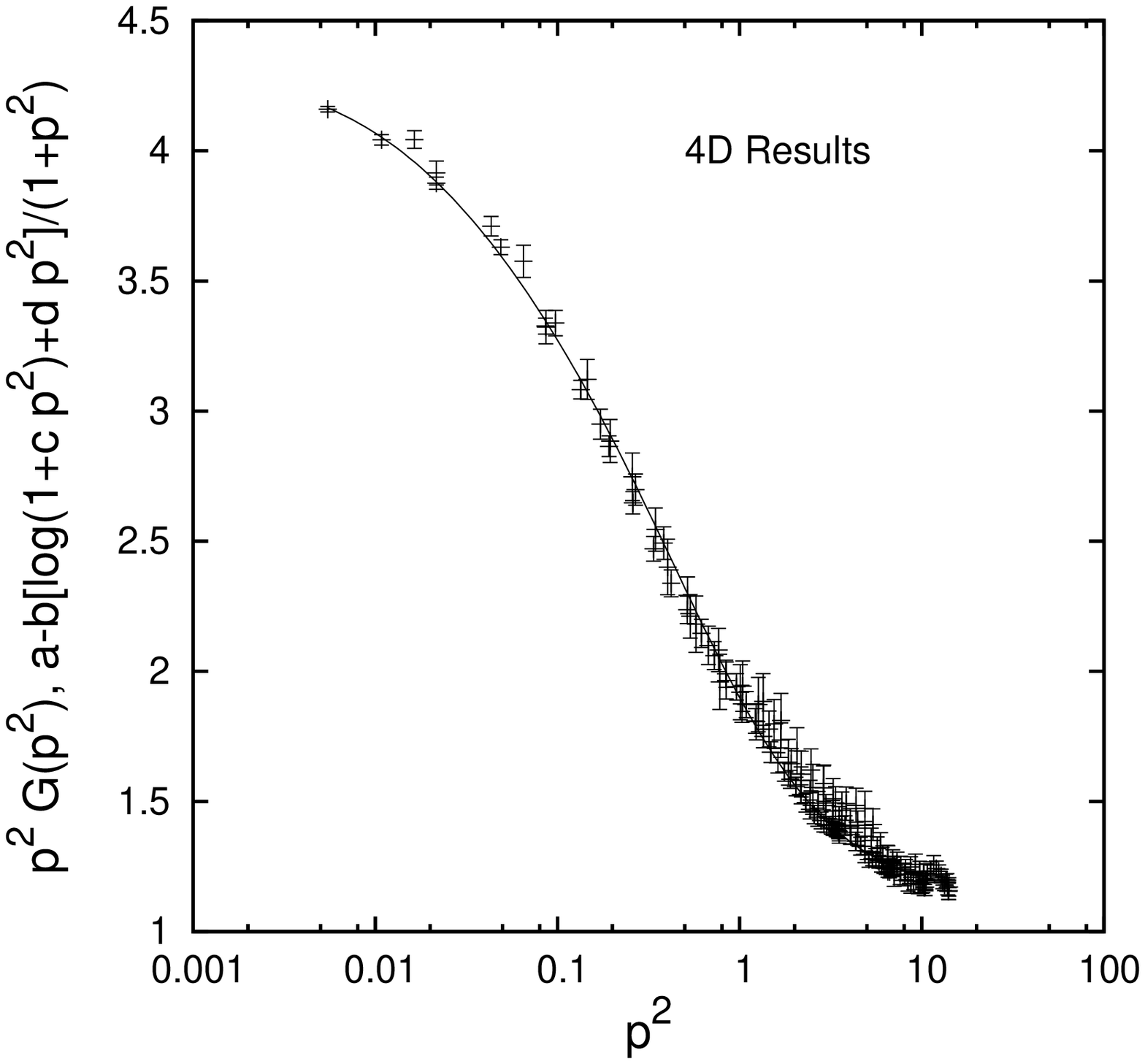}
\hspace{-1mm}
\includegraphics[scale=0.43]{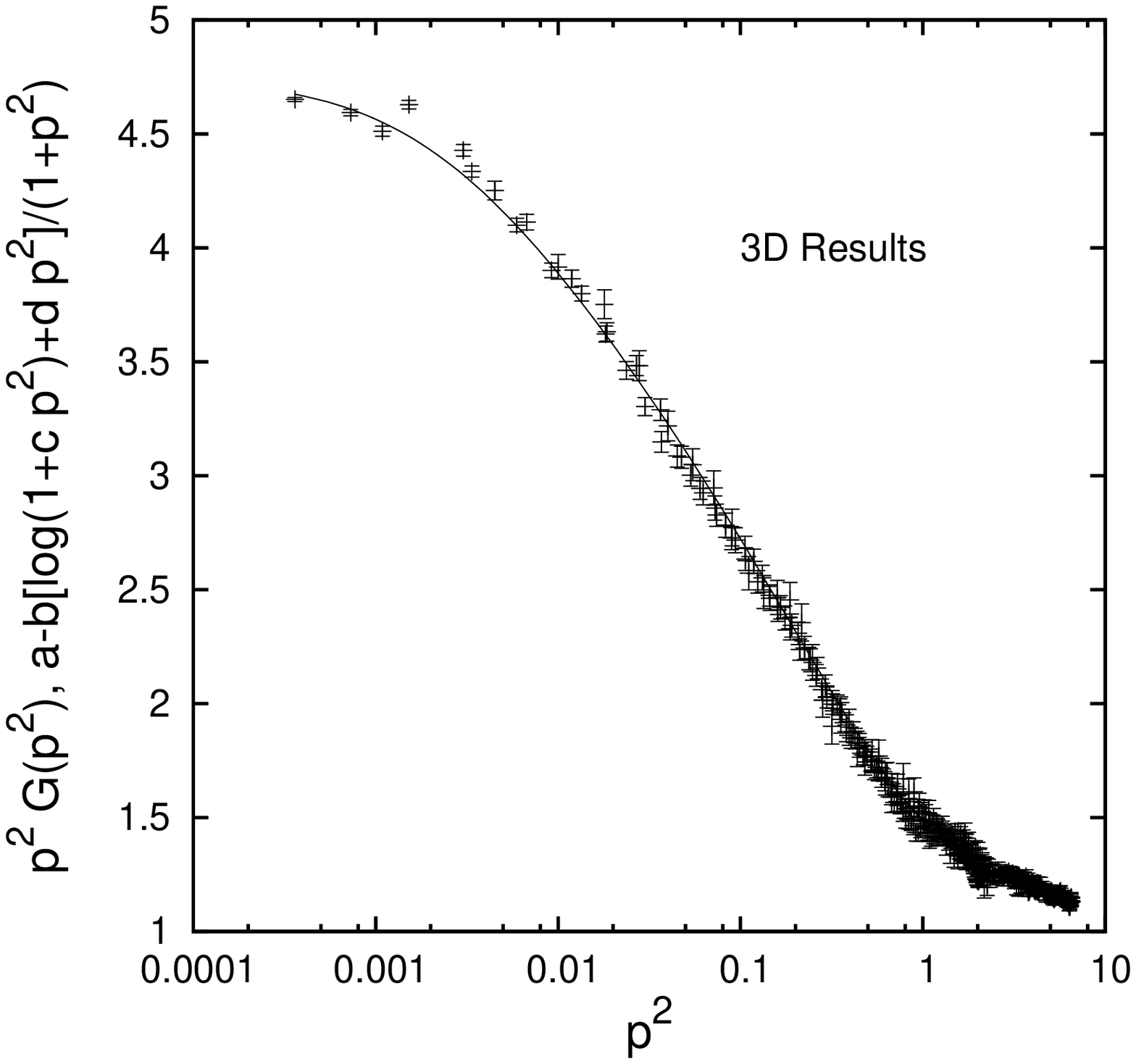}
\vspace{-26mm}
\caption{
  The ghost dressing function $p^2 G(p^2)$ as a function of the 
  lattice momenta $p^2$ (both in physical units) for the pure-$SU(2)$ 
  case in $d=4$ (left), considering volumes of up to $128^4$
  (lattice extent $\sim 27$ fm), and $d=3$ (right),
  considering volumes of up to $320^3$
  (lattice extent $\sim 85$ fm). Note the logarithmic scale for $p^2$.
  We also show the fits to 
  $a - b \left[ \log(1 + c p^2) + d p^2 \right] / (1 + p^2)$.
  For $d=4$, the fit parameters $a,b,c,d$ are respectively 
  4.32(2), 0.38(1), 80(10), 8.2(3).
  }
\label{ghost}
\end{figure}

Our data for the ghost propagator support a tree-level (or free) form
in the infrared limit. This behavior is better seen if one considers 
the dressing function $p^2 G(p^2)$, as shown in Fig.\ \ref{ghost}. 
Indeed, the data can be well fitted \cite{Cucchieri:2008fc}
by the form $a - b \left[ \log(1 + c p^2) + d p^2 \right] / (1 + p^2)$,
consistent with $\kappa = 0$ in the infrared limit.
Note also that, for smaller $p$, this form is equivalent to
$a - b \log(1 + c p^2)$ (proposed in \cite{Aguilar:2008xm}), where
$c$ may be related to a gluon mass. This is also observed in $d=3$. 
We remark again that enhancement is seen at intermediate momenta
and that, depending on how the fit parameters change with the lattice
spacing, there might be a logarithmic enhancement in the continuum limit.

\section{$\beta=0$}
\label{beta0}

An interesting laboratory to test for various sources of systematic errors 
in the simulations is the apparently trivial case of $\beta=0$, i.e.\ 
no dynamics from the lattice Boltzmann weight associated with the action
[see Eq.\ (\ref{eq:Ug})]. This was considered in 
\cite{Cucchieri:1997fy,beta0,Cucchieri:2009zt}.
%\cite{Cucchieri:1997fy,Sternbeck:2008mv,Cucchieri:2009zt,Maas:2009ph}.
Since $\beta=0$ corresponds to an unphysical limit, the issue of setting 
the lattice scale (given by $a$) is delicate. As discussed in
detail in \cite{Cucchieri:2009zt}, a possible choice is $a\to\infty$.
This is convenient because the lattice extent will already 
be infinite and there will be no finite-size effects. On the other hand,
if $a$ is large and $p$ is not very small there could be discretization 
effects. We must see from 
the data which effect is predominant. (The latter effect may also be 
measured in terms of breaking of rotational invariance.)
As shown in \cite{Cucchieri:2009zt}, we see: 1) clear violation of
reflection positivity for the gluon propagator, 2) a seemingly
finite and nonzero limit for $D(0)$, including analysis with the 
bounds in Eq.\ (\ref{eq:Dbounds}), 3) a very good fit of $p^2 G(p^2)$ 
to the form $a - b \log(1 + c p^2)$ and 4) no finite-volume or 
rotational-symmetry-breaking effects, suggesting that the data are
in the deep infrared limit and at infinite volume.
Essentially, we see the same infrared behavior as for finite $\beta$.
Note that our conclusions differ somewhat from \cite{beta0}.
% \cite{Sternbeck:2008mv,Maas:2009ph}.

\section{Conclusion}
\label{conclusion}

Lattice simulations of infrared Landau gluon and ghost propagators have
come a long way in the past couple of years. 
The current paradigm is that of a massive gluon and a free ghost,
as proposed long ago by some Dyson-Schwinger-equation studies
(see e.g.\ \cite{DS}).
% (see e.g.\ \cite{Cornwall:1981zr,Aguilar:2004sw,Boucaud:2008ky}).
These simulations are greatly influenced by interchange with 
researchers who use analytic and semi-analytic methods, and vice-versa.
This synergy has contributed to making infrared QCD a very productive 
field, with an active community and several dedicated workshops every year.

\vskip 6mm
\noindent
{\bf Acknowledgements}:
The authors thank the organizers of QCD-TNT09 for a very stimulating 
meeting and acknowledge partial support from the Brazilian Funding Agencies 
FAPESP and CNPq. T.M. also thanks the Theory Group at DESY-Zeuthen for 
hospitality and the Alexander von Humboldt Foundation for financial support.

\end{document}